\documentclass[aps,prb,twocolumn,superscriptaddress,showpacs,floatfix,10pt]{revtex4-2}

\usepackage{amsmath}
\usepackage{amsfonts}
\usepackage{graphicx}
\usepackage{color}
\usepackage{epstopdf}
\usepackage{natbib}
\usepackage{xfrac}
\usepackage{enumitem}
\usepackage[english]{babel}
\usepackage{booktabs}

\usepackage{float} 
\usepackage{subfigure}

\begin{document}
\title{Transfer Matrix description of heterostructured spintronics Terahertz emission}

\author{Yingshu~{Yang}}
\email{yingshu001@e.ntu.edu.sg}
\affiliation{School of Physical and Mathematical Sciences, Nanyang Technological University, Singapore, Singapore}

\author{Stefano~{Dal~Forno}}
\email{sdalforno@ntu.edu.sg}
\affiliation{School of Physical and Mathematical Sciences, Nanyang Technological University, Singapore, Singapore}

\author{Marco~{Battiato}}
\email{marco.battiato@ntu.edu.sg}
\affiliation{School of Physical and Mathematical Sciences, Nanyang Technological University, Singapore, Singapore}

\date{\today}

\begin{abstract}

In this work we developed an extension to the Transfer Matrix method (TMM) to include a current source term, with the aim of describing both the transmission and emission of THz pulses in spintronics THz emitters. The TMM with source is derived from the Maxwell equations with a volume free current term.
This extension to the TMM allows for the study of realistic spintronics THz emitters as multilayers of different materials with different thicknesses. 
We use this to prove that, in spite of a common misconception, the drop rate of the THz emission amplitude at higher heavy metal layer thicknesses is not related and does not provide information about the spin diffusion length.  It is instead the effect of the increase in the parasitic  absorption of the generated radiation by the conducting parts of the spintronics THz emitter itself.

\end{abstract}


\maketitle

The terahertz (THz) frequency gap is a frequency range between infrared and microwaves (0.1--10 THz).
This frequency range is drawing much attention in a number of fields aimed at the fabrication of new optoelectronic devices.
Examples of the applicability of THz technology are security checks\cite{Michael_THzsecurity_2003}, nondestructive biomedical testings\cite{siegel_THztesting_2004}, communications\cite{federici_THzreview_2010,song_THzpresent_2011} and time-domain spectroscopy (THz--TDS)\cite{neu_tutorial_2018}.
For these reasons it is important to build stable, cheap and performant THz emitters.
At present, the  most commonly used THz emitters are made of photoconductive antennas \cite{mourou_picosecond_1981,lepeshov_enhancement_2017,tani_photoconductive_2012}, quantum cascade lasers \cite{kohler_terahertz_2002,burghoff_terahertz_2014} and electro-optic crystals\cite{hebling_velocity_2002,hu_free_1990}.
These emitters are able to cover different ranges of THz frequencies, however, they struggle to cover frequencies spanning from 0.1THz to 10THz \cite{chen_current_2019}.

For the past few years a new type of emitter, known as spintronics THz emitters, has attracted a great interest in the field by providing low-cost, broad spectrum, flexible and performant emission based on spin-to-charge conversion in ferromagnet(FM)/heavy metal(HM) multilayers \cite{kampfrath_terahertz_2013,Seifert2016SpintronicsTHzemitters,torosyan_optimized_2018,feng_highly_2018,herapath_impact_2019,nandi_antenna-coupled_2019}. 
In this kind of THz emitters an ultrafast spin current pulse is generated inside the ferromagnetic (FM) layer due to superdiffusive spin transport triggered by an optical pulse \cite{battiato2010superdiffusive,battiato2012superdiffusive,battiato2014superdiffusive,battiato2016superdiffusive,battiato2017superdiffusive}. 
Subsequently, the spin current flows into the heavy metal layer and, due to the inverse Spin Hall effect (ISHE), it is converted into a transverse charge current which in turn produces THz radiation \cite{valenzuela_direct_2006,saitoh_conversion_2006}.

\begin{figure}[htb]
\centering
\includegraphics[width=\linewidth]{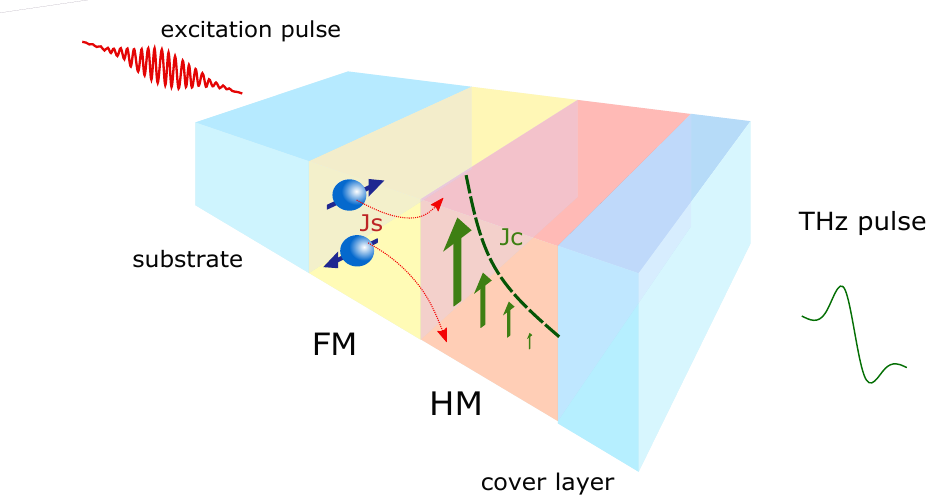}
\caption{A schematic description of the working principle of the Spintronics THz Emitter and their geometry.}
\label{fig:structure}
\end{figure}

The effort in optimising spintronics THz emitter requires the ability to accurately describe these processes.
An effective approach to model the propagation of electromagnetic waves in heterostructure devices is the transfer matrix method (TMM) as it efficiently and elegantly accounts for multiple reflections at the different interfaces of a multilayer. 
The standard TMM however addresses the problem of a wave crossing a multilayer, while in the case of the spintronics THz emitters, the wave is generated within one (or more) of the layers by a pulsed current. 
Previously, there have been several attempts at modifying the TMM by approximating the current as a 2D current running through the interface between two of the layers\cite{cheng2019spininjectionsemiconductors} or by considering the interface as a conducting interface\cite{khorasani_modified_2002} so as to address specific applications such as integrated optical modulators\cite{biswas_integro-differential_2000}, memory, transistors\cite{khorasani_new_2001} and programmable diffractive elements\cite{rashidian2001new}.
However, to describe spintronics THz emission, these treatments do not address the experimental setup and either oversimplify the geometry, assume too idealised conditions, or are not sufficiently general. 
In spintronics THz emitters, the generated charge current is not necessarily only a 2D current, but extend over the entire HM layer (and often more than the HM layer) as show in Fig.~\ref{fig:structure}. 
Therefore, it is important to describe the THz emission in a more realistic way.

In this work we developed an extension to the TMM to include space- and time-resolved charge currents $J[t,z]$. 
The approach can be straightforwardly adapted to the case of free charge distributions, yet we do not address this case here.
Specifically, we focused on photo induced ultrafast spin currents
\cite{battiato2010superdiffusive, battiato2012superdiffusive, battiato2014superdiffusive, battiato2016superdiffusive, battiato2017superdiffusive}
via the inverse spin Hall effect \cite{kampfrath_terahertz_2013,Seifert2016SpintronicsTHzemitters}, yet the approach below is valid independently on the physical origin of the free current or charge.
By enforcing the appropriate boundary conditions, we propagated the generated wave through the multilayer system.
This provides an accurate description of the THz emission and transmission through the whole structure.

\section{Maxwell equations with source}
We first focus on a single layer and write the free current in terms of its Fourier components in space and time
\begin{equation} \label{EQfouriercurrent}
 	J[t,z]= \sum_{\omega,k}J [k,\omega] \,\exp[i(k z-\omega t)]
\end{equation}
where $i$ is the imaginary unit.
As the equations we are going to address are linear, we can explicitly compute the emitted radiation for the case of a single Fourier component and then  obtain the full solution by summation.
Therefore we study the case of a source $J[z,t]=J \exp[i(k z-\omega t)]$. The Maxwell's equations within the layer where the free current is running are
\begin{align}
	\partial_z E\left[t,z\right]=&\mu\, \partial_t H\left[t,z\right] \\
	\partial_z H\left[t,z\right]=&\epsilon\, \partial_t E\left[t,z\right] + J e^{i(k z-\omega t)}
\end{align}
where we stress that, in general, $k\ne \omega \sqrt{\epsilon \mu}$, as $k$ is the reciprocal space variable (momentum) for the spatial Fourier transform of the current profile and not the light wave vector. Notice how $k$ is a purely real number, while $\omega \sqrt{\epsilon \mu}$ is, in general, a complex value.

The solution of the inhomogeneous system  above can be constructed  as the general solution of the associated homogeneous system plus a particular solution of the inhomogeneous system. The general solution is known from textbooks \cite{born_principles_2013}  and a summary of the approach that uses a notation consistent with the present work can be found in the introductory sections of Ref.~\onlinecite{yang_removal_2020}.
We look for a particular solution of plane waves as in Eq.~\ref{EQfouriercurrent}. By substituting inside the inhomogeneous system we obtain the amplitudes of the fields for the particular solution,
\begin{align}
	E_p = J \frac{-i\omega \mu}{\epsilon \mu \omega^2-k^2}\\
	H_p = J \frac{ik}{\epsilon \mu \omega^2-k^2} .
\end{align}
Therefore, the time- (notice not space-) Fourier transform of general solution can be written as
\begin{equation} \label{EQFields}
    	\bar{F}\left[\omega,z\right] = \begin{bmatrix} E\left[\omega,z\right]\\ H\left[\omega,z\right]\end{bmatrix} 
    	=  \bar{\bar{a}}\left[ \omega,z \right] \begin{bmatrix} f^{>} \\ f^< \end{bmatrix} + J \begin{bmatrix} \frac{-i\omega \mu e^{ikz}}{\epsilon \mu \omega^2-k^2} \\  \frac{ik e^{ikz}}{\epsilon \mu \omega^2-k^2} \end{bmatrix},
\end{equation} 
with 
\begin{align}
    &\bar{\bar{a}}\left[ \omega,z \right]=\begin{bmatrix} e^{i \omega \sqrt{\epsilon \mu} z } &e^{-i \omega \sqrt{\epsilon \mu} z} \\ 
    -\sqrt{\frac{\epsilon}{\mu}}e^{i \omega \sqrt{\epsilon \mu} z }&{\sqrt{\frac{\epsilon}{\mu}} e^{-i \omega \sqrt{\epsilon \mu} z}}\end{bmatrix},
\end{align}
where $f^{>}[\omega]$ and and $f^{<}[\omega]$  are two amplitude parameters that are to be fixed by boundary conditions and that represent the amplitude of the right propagating wave and of the left propagating one, respectively.
For shortness we can rewrite Eq.~\ref{EQFields} as
\begin{equation}
	\bar{F}\left[\omega,z\right]=\bar{\bar{a}}\left[ \omega,z \right]\bar{f} + J\, \bar{b}[\omega,k,z],
\end{equation}
with
\begin{equation}
	\bar{b}[\omega,k,z]=\frac{i \,e^{i kz}}{\epsilon \mu \omega^2-k^2}  \begin{bmatrix} -\omega \mu\\ k\end{bmatrix},
\end{equation}
which gives the fields amplitude at a given position.
The shape of the fields within layers that have no current running through them is simply a special case of Eq.~\ref{EQFields} with the free current term set to zero.
\section{Transfer matrix with source}

We now suppose we have a multilayer with $M$ layers, sandwiched by air. We assume that the $N$-th layer has a free current at a given frequency and momentum. 
We want to calculate the radiation generated by the free current yet including transmission and multiple reflections through the multilayer.
We compute the response at the same frequency as the source. The fields at the interfaces of the $N$-th layer are given by
\begin{align}
	\bar{F}_N\left[0\right]&=\bar{\bar{a}}_N\left[ 0 \right]\bar{f}_N + J_N[k]\, \bar{b}_N[k,0]\\
	\bar{F}_N\left[d_N\right]&=\bar{\bar{a}}_N\left[ d_N \right]\bar{f}_N+ J_N[k]\, \bar{b}_N[k,d_N].
\end{align}
where we have used the subscript $N$ to refer to quantities that depend on the characteristics of the $N$-th layer. For compactness, we have dropped the explicit dependence on the frequency in the above expressions.  
Since both  $\bar{F}_N\left[0\right]$ and $\bar{F}_N\left[d_N\right]$ depend on the amplitude parameters $\bar{f}_N$, they can be  written in terms of each other:
\begin{equation} \label{EQfieldssourcelayerinterfaces}
\begin{split}
	\bar{F}_N\left[d_N\right]= & \bar{\bar{M}}_N\left[ d_N \right]\bar{F}_N\left[0\right] + \\
	& + J_N[k] \left(\bar{b}[k,d_N] - \bar{\bar{M}}_N\left[ d_N \right] \bar{b}_N[k,0]\right),
\end{split}
\end{equation}
where,
\begin{equation}
    \bar{\bar{M}}_N[d]=\bar{\bar{a}}_N\left[d \right]\left(\bar{\bar{a}}_N\left[0 \right]\right)^{-1}.   
\end{equation}
Eq.~\ref{EQfieldssourcelayerinterfaces} provides the amplitude of the fields at the right interface of the $N$-th layer if the fields on the left interface are known.

A similar approach can be used for all the other layers. As we assume here that no current runs in any of the other layers, except the $N$-th one (the approach can be easily generalised), for a generic layer $n$ we can still use Eq.~\ref{EQfieldssourcelayerinterfaces} with the source term set to zero:
\begin{equation} 
	\bar{F}_n\left[d_n\right]=\bar{\bar{M}}_n\left[ d_n \right]\bar{F}_n\left[0\right].
\end{equation}
Using the fact that the fields must be continuous at every interface, we can obtain (by a procedure that very closely resembles the transfer matrix method derivation\cite{born_principles_2013,yang_removal_2020}) the relationship between field amplitude parameters of the air on the left of the multilayer ($\bar{f}_{0}$) and those of the air on the right ($\bar{f}_{\infty}$)
\begin{align}
	\bar{f}_{\infty}=&\; \bar{\bar{T}}_{[0,\infty]} \bar{f}_{0} + J_N[k] \,\bar{\bar{T}}_{[N,\infty]} \cdot  \\
	&\cdot \left(\left( \bar{\bar{a}}_N\left[d_N \right]\right)^{-1} \bar{b}_N[k,d_N]- \left( \bar{\bar{a}}_N\left[0 \right]\right)^{-1} \bar{b}_N[k,0]\right), \nonumber
\end{align}
or for brevity
\begin{equation}
	\bar{f}_{\infty}= \bar{\bar{T}}_{[0,\infty]} \bar{f}_{0} + \begin{bmatrix} J^> \\ J^<\end{bmatrix} ,
\end{equation}
where
\begin{equation}
    \bar{\bar{T}}_{[n,m]}=\left(\bar{\bar{a}}_{m}\left[0 \right]\right)^{-1} \left(\prod_{j=m-1}^{n+1} \bar{\bar{M}}_j[d_j] \right)\bar{\bar{a}}_n\left[0 \right],    
\end{equation}
with $n$ and $m$ representing two layers (the reader should notice the inverse order of the product).

We then obtain the emitted radiation by solving the system of equations,
\begin{equation} \label{EQonlyemission}
	\begin{bmatrix} f_{\infty}^>\\ 0\end{bmatrix} = \bar{\bar{T}}_{[0,\infty]} \begin{bmatrix} 0\\  f_{0}^<\end{bmatrix} + \begin{bmatrix} J^> \\ J^<\end{bmatrix} ,
\end{equation}
where  $f_{\infty}^>$ gives the amplitude of the radiation emitted towards the right of the multilayer, while  $f_{0}^<$ the one towards the left. Eq.~\ref{EQonlyemission} specifies that there is no radiation incoming towards the multilayer on either side.
The system can be solved and, in particular, the righthand side emission is,
\begin{equation}
	f_{\infty}^> = J^>-\frac{T_{[0,\infty],12}}{T_{[0,\infty],22}}J^<,
\end{equation}
where the numerical indices represent matrix elements corresponding to a specific column and row.

\subsection{Echo and probe pulse}

It is useful to generalise the treatment above to include two further effects: the presence of an external (probe) pulse for which we need to calculate the transmission, and the removal of echo pulses eventually produced in a thick substrate from the transmitted and generated radiation spectrum. 

The first situation is common for THz pump and probe experiments. The pump pulse triggers the formation of the pulsed current (as in the case of spintronics THz emitters) that, in turns, creates a pulsed emitted radiation which is experimentally measured. However it is common to further investigate the system by sending a  probe pulse onto the multilayer, and investigating its transmission through or reflection from the sample.

The second effect leads to some complications in the presence of echoes and we addressed it in Ref.~\onlinecite{yang_removal_2020}. Without entering too much into details, if one of the layers is particularly thick (as usually the substrate is), the radiation can generate echo pulses which leave the system with a relatively large time delay. Using time domain spectroscopy, such pulses are seldom measured, since only a finite time window is experimentally accessed. Therefore only part of the emitted or transmitted time domain wave is measured and subsequently used to construct the experimental spectrum: echo pulses are excluded. The transfer matrix method, on the other hand, generates the full signal including echo. We have showed that the two spectra are importantly different and how to explicitly theoretically construct the spectrum with the echo removed.\cite{yang_removal_2020}

Instead of using Eq.~\ref{EQonlyemission}, we can include both the effects mentioned above by solving (see Ref.~\onlinecite{yang_removal_2020} for details)
\begin{align}
	&\bar{f}_{\infty}=\bar{\bar{T}}_{[S,\infty]} \begin{bmatrix} t_{\left[0,S\right]} f_{0}^>\\  f_{S}^<\end{bmatrix} +  \\
	&+J\, \bar{\bar{T}}_{[N,\infty]} \left(\left( \bar{\bar{a}}_N\left[d_N \right]\right)^{-1} \bar{b}[k,d_N]- \left( \bar{\bar{a}}_N\left[0 \right]\right)^{-1} \bar{b}[k,0]\right) \nonumber.
\end{align}
where the index $S$ refers to the substrate layer, and $f_{0}^>$ represents the probe pulse sent from the lefthand side of the multilayer.
This gives 
\begin{equation}
	f_{\infty}^> = t_{[0,S]}\;t_{[S,\infty]} f_{0}^> + J^>-\frac{T_{[S,\infty],12}}{T_{[S,\infty],22}}J^<,
\end{equation}
where $t_{[n,m]}$ represents the transmission coefficient,
\begin{equation}
    t_{[n,m]} = \frac{f^{>}_{m}}{f^{>}_{n}} = \frac{T_{[n,m],11}T_{[n,m],22}-T_{[n,m],12}T_{[n,m],21}}{T_{[n,m],22}}.
\end{equation}

\subsection{Propagation, focusing and detector's response}
The field amplitudes computed above are the ones immediately outside of the multilayer. However, the usual time-domain terahertz spectroscopic setups do not measure those fields directly.
Typically, the terahertz radiation, before reaching the detector propagates in air and is directed through parabolic mirrors to focus on the electro-optic detection crystal. 

The propagation and focusing process can be approximately included by multiplying the transmission by a response function that is linear with frequency \cite{kampfrath_terahertz_2013}.

However, the electro-optic detector response results from several effects within the electro-optic detecting crystal itself. Thus, we avoided the details of the THz detection process since it varies with different setups and is beyond the scope of this work. But it is important to note that the response from electro-optical crystal further convolute with the emitted terahertz electric field before its final detection.

\section{Spintronics THz emitters: optical pump - THz probe}

We use the approach developed above to model a typical optical pump - THz probe setup for spintronics THz emitters. We choose, as an example, a multilayer system commonly used for spintronics THz emitters: a 1mm $Al_2O_3$ substrate, a 10nm $Fe$ layer, a 2nm $Au$ layer, and a 10nm $Al_2O_3$ capping layer.

The dielectric response for each layer is fully accounted for as frequency dependent. The frequency-dependent relative permittivities for $Fe$ and $Au$ are constructed using the Drude model with plasma frequency and damping frequency taken as  4.08eV and 0.02641eV respectively for $Fe$, and as 9.03eV and 0.027eV for $Au$~\cite{Ordal_constants_1988,ordal_constants_1987}. The dielectric response of sapphire has been taken as constant within the THz range and given by  $\epsilon_r = (n+i\kappa)^2$, where $n = 3.31 $ and $\kappa = 0.002 $ \cite{sanjuan_optical_2012,palik_handbook_1998,kitamura_optical_2007}. We remind that the method straightforwardly allows for the use of more general frequency dependencies, as for instance permittivities interpolated from experimental data.

\begin{figure}[tb]
\centering
\includegraphics[width=\linewidth]{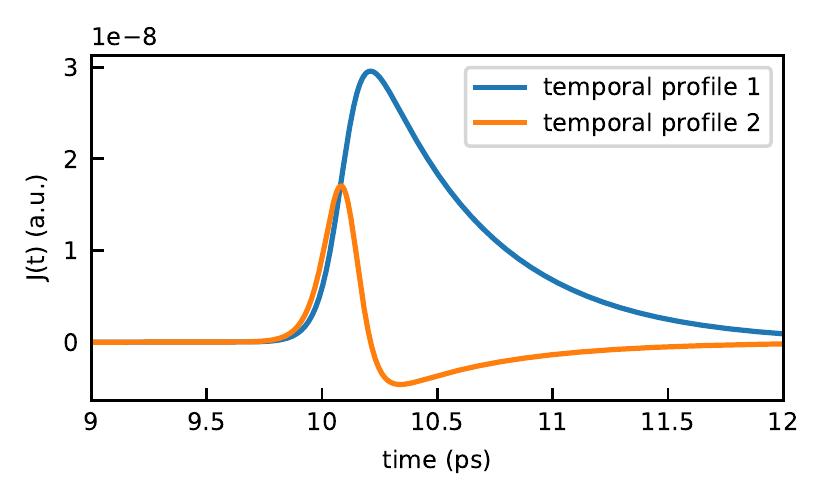}
\caption{The two different temporal profiles for the free current, used in the simulation. }
\label{fig:timeprofiles}
\end{figure}

We assume, in this example, that the transverse charge current is localised within the $Au$ layer only (but more general configurations can be described as well). We construct the temporal and spatial profile of the charge current as the product of a spatial and a temporal function. We test our approach with two temporal profiles for the charge current. The first used profile is
\begin{equation}     J_{1st}(t)= h\; \frac{e^{\frac{(t-t_0)-\sfrac{a}{2}}{\sfrac{a}{4}}}}{e^{\frac{(t-t_0)-\sfrac{a}{2}}{\sfrac{a}{4}}}+1} \;e^{-\frac{t-t_0}{b}},
    \label{fun:timeprofile1}
\end{equation}
with a rise time of $a = 200$fs and a switching-off time of $b = 500$fs (see temporal profile 1 in Fig.~\ref{fig:timeprofiles}). Here $h$ refers to the approximate peak value and taken to be $5 \times 10^{-8}$, while $t_0$ gives the approximate time position of the temporal profile and is taken as 10ps. The second profile is meant to resemble the experimental and theoretical ones in Ref.~\onlinecite{kampfrath_terahertz_2013}, and we construct it here as the time derivative of the first time profile (see temporal profile 2 in Fig.~\ref{fig:timeprofiles}):
\begin{align} 
    J_{2nd}(t)&=\frac{J_{1st}(t)}{dt}.
    \label{fun:timeprofile2}
\end{align}
To produce a comparable amplitude to the 1st profile the $h$ in Eq.~\ref{fun:timeprofile2} is taken as $5 \times 10^{-21}$.

\begin{figure}[tb]
\centering
\includegraphics[width=\linewidth]{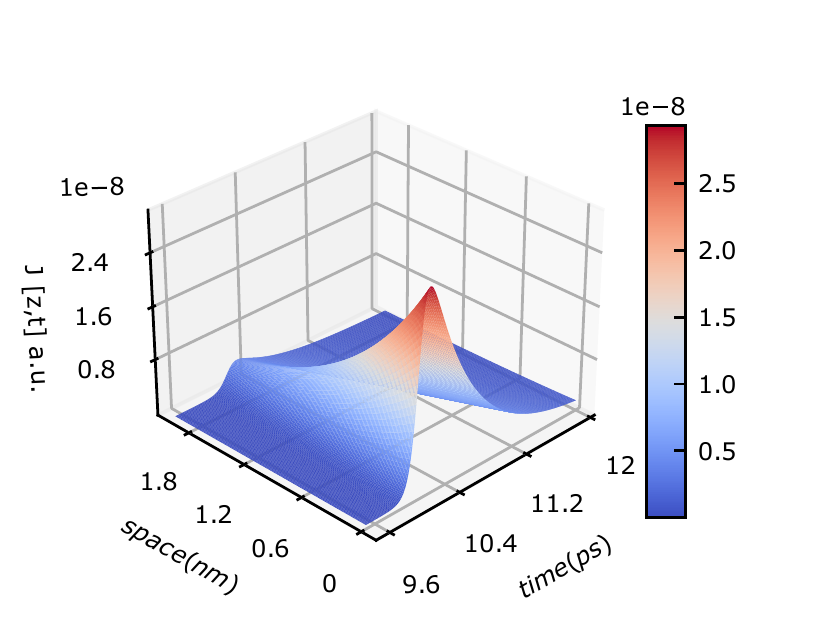}
\caption{An example of the 3D plotting of the temporal (profile type 1) and spatial profile of the source current density.}
\label{fig:3Dprofile}
\end{figure}

As spatial profile we assume a simple exponential decay with depth in the $Au$ layer:
\begin{equation}
    J(z) = e^{-\frac{z}{\lambda}}.
    \label{fun:spatialprofile}
\end{equation}
In principle the superdiffusive spin current is not expected to have such profile, especially in such thin layers. Nonetheless due to the very high spin-orbit coupling of the $Au$ layer, spin to charge conversion is expected to quickly reduce the spin polarisation of the spin current. Therefore, in high spin orbit coupling materials, we expect the charge current induced by inverse spin Hall effect to quickly drop, not because of the drop in amplitude of the superdiffusion of the excited carriers, but due to the quick loss in their spin polarisation. For simplicity, we assume here a decay length (spin diffusion length) $\lambda$ of $1$ nm (see Fig.~\ref{fig:3Dprofile}). 
It is however beyond the scope of the present work to investigate the spatial profile of the spin-to-charge conversion mechanism, and the profile above is to be intended mostly as a demonstrative example.

\subsection{Emission and THz probe}

\begin{figure}[tb]
\centering
\includegraphics[width=\linewidth]{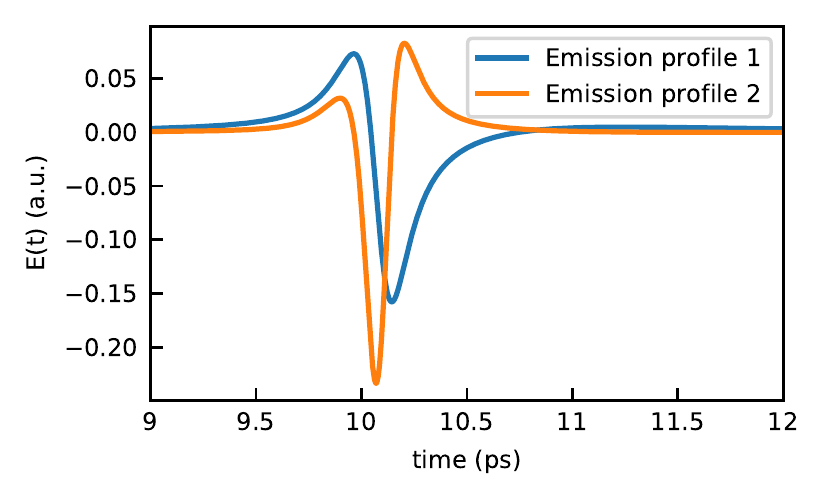}
\caption{The emitted THz wave using different temporal profiles. Emission profile 1 comes from temporal profile 1. Emission profile 2 comes from temporal profile 2. }
\label{fig:emission_profile_both}
\end{figure}

Using the TMM with source developed above, we are able to compute the temporal profile emitted THz wave using the two types of currents in Eq.~\ref{fun:timeprofile1} and \ref{fun:timeprofile2} as shown in Fig.~\ref{fig:emission_profile_both}. 
These waveforms carry information about the temporal profile of the source current, but also about the materials that compose the multilayer, and any other effect happening during propagation and detection.

\begin{figure}[tb]
\centering
\includegraphics[width=\linewidth]{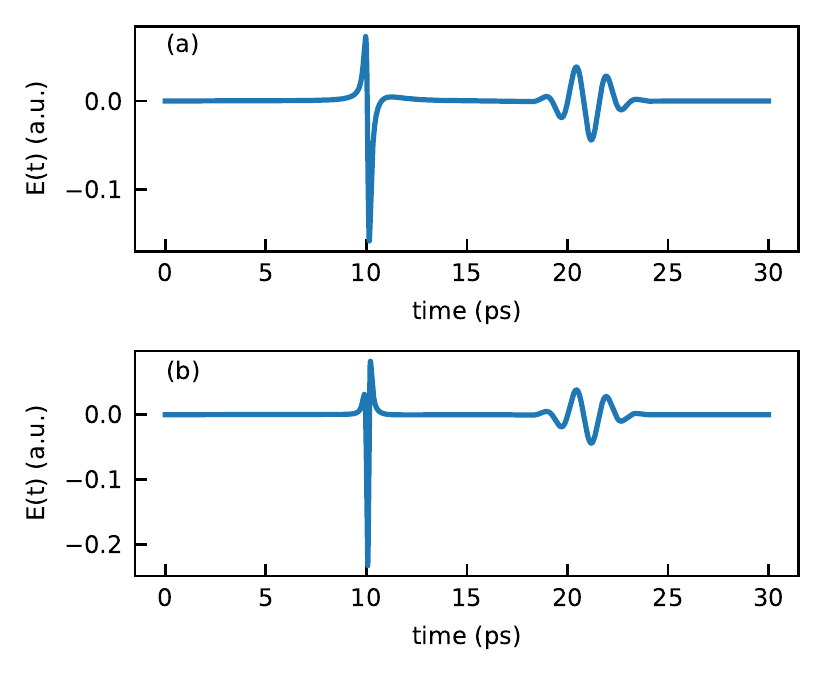}
\caption{The emitted THz signal together with the transmitted THz probe. (a) Emitted profile 1 with transmitted probe. (b) Emitted profile 2 with transmitted probe. }
\label{fig:probeandemission}
\end{figure}

Our method is also able to reproduce the transmitted THz probe and the emitted THz signal within the same formalism (see Fig.~\ref{fig:probeandemission}). 

In the computed case, both the source current and the probe signal are generated around time 10ps. The emitted pulse however reaches the surface on the right very quickly, while the probe signal is still travelling through the substrate (see Fig~\ref{fig:structure}). For that reason, the two pulses are observed with a large time delay. The model can however describe also situations where the two pulses overlap in time.

\subsection{Effect of HM thickness on THz amplitude}

We are now ready to address the core interesting point. It might be tempting to assume that making the HM layer thicker should increase the THz emission efficiency, if the spin diffusion length is long enough. The rationale is that a thicker HM layer will allow for more charge current to be generated. That concept, although partially correct, fails to include another important effect of the increase in thickness of the HM layer. This layer is (in standard spintronics THz emitters) conducting. THz radiation is heavily absorbed by conducting layers. Therefore, while a thicker HM increases the amount of current, it also increases the absorption of the generated THz. We investigate here the overall balance of these two competing effects.

\begin{figure}[tb]
\centering
\includegraphics[width=\linewidth]{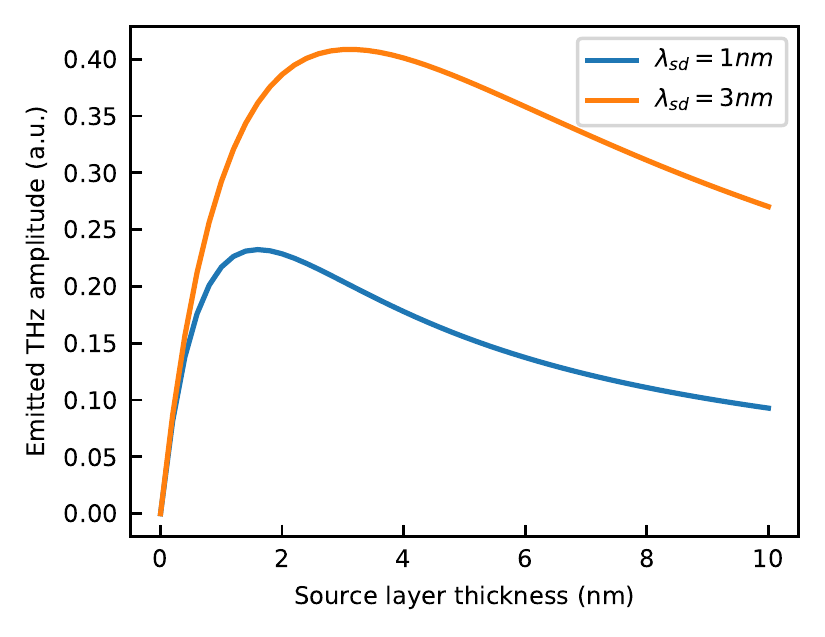}
\caption{Dependence of THz peak to peak amplitude on the thickness of source layer, computed for two different spin diffusion lengths.} 
\label{fig:AmpChange}
\end{figure}

We assume two different scenarios: one with a spin diffusion length of 1nm and one with a spin diffusion length of 3nm. We use a sample similar to the one used above, but with an HM layer with increasing thickness:  $Al_2O_3$ (1mm, substrate) / $Fe$ (10nm)/  $Au$ ($X$nm)/  $Al_2O_3$ (10nm), where $X$ goes from 0 to 10.
As expected, we observe in Fig.~\ref{fig:AmpChange} that the larger the spin diffusion length, the higher the signal is. 
This is a trivial consequence of the fact that the spatial integral of the source current is larger the longer the spin diffusion length is. However, we can also observe that the decay rate of the emission efficiency at higher thicknesses  is not indicative of the spin diffusion length. It is rather dependent on the increase of absorption in the conducting HM layer. 
The spin diffusion length however impacts the position at which the efficiency vs thickness reaches its maximum in our simulations. 
We however warn against using this conclusion in realistic cases. One important assumption here was that both the spin diffusion length and the current spatial profile would not be affected by the HM layer thickness. 
The spatial profile was, in our case, an exponential decrease with depth. This is however the result of a very simplistic choice for the spatial profile and is unlikely to be the case in real cases. 
In those cases, a proper description and parametrisation of the spin superdiffusion profile, spin-to-charge conversion, and material properties is required. In that case, the thickness at which the maximum is observed is going to be indicative of a more complex interplay between the shape of the multilayer and the spin transport.

Finally we mention that similar results are obtained by changing the thickness of the FM layer (not shown), due to very similar reasons.

\section{Conclusion}

In conclusion, we have proposed an analytical way to describe the spintronics THz emitter based on an extension of the Transfer Matrix Method to include volume sources (beyond the previously used approximations of 2D currents). The TMM with source is flexible, easy to implement, performant and able to describe a large variety of spintronic THz emitters constructed by different materials. It also provides a detailed prediction of the emitted THz waveform generated by imposing specific spatial and temporal profiles for the source current. It moreover easily allows to simulate both a transmitted THz probe and generated THz pulse at the same time.

Using this approach we have been able to address the very important issue of how the thickness of the layers influences the efficiency of the THz emission and its relationship to the spin diffusion length. We have proven that, even if it is tempting to do so, the drop rate of the THz emission at higher thicknesses is not related and does not provide information about the spin diffusion length, but it is the effect of the increase in the parasitic  absorption of the generated radiation by the conducting parts of the spintronics THz emitter itself.

\bibliography{main}

\end{document}